# Toward an atlas of the number of visible stars

Pierantonio Cinzano [a], Fabio Falchi [a,b,*]

[a] Istituto di Scienza e Tecnologia dell'Inquinamento Luminoso (Light Pollution Science and Technology Institute), 36016 Thiene, Italy
[b] Departamento de Física Aplicada, Universidade de Santiago de Compostela, 15782 Santiago de Compostela, Galicia (Spain)

Modelling techniques for the propagation of light pollution in the atmosphere allow the computation of maps of artificial night sky brightness in any direction of the sky, involving a large number of details from satellite data. Cinzano et al. (2001a) introduced a method of mapping naked eye star visibility at the zenith from large areas based on satellite radiance measurements and Garstang models of the propagation of light pollution. It takes into account the altitude of each land area from digital elevation data, natural sky brightness in the chosen sky direction based on the Garstang approach, eye capability after Garstang and Schaefer, and atmospheric extinction in the visual photometric band. Here we discuss how to use these methods to obtain maps of the average number of visible stars when looking at the night sky hemisphere, finally answering, site by site, the question of how many stars are visible in the sky. This is not trivial, as the number of stars visible depends on the limiting magnitude in each direction in the sky, and this depends on sky brightness in that direction, atmospheric extinction at that zenith distance and the observer's visual acuity and experience. We present, as an example, a map of the number of visible stars in Italy to an average observer on clear nights with a resolution of approximately 1 km.

atmospheric effects – site testing – scattering – light pollution

## 1 Introduction

The state of the night sky and light pollution are arguments which frequently face the general public due to the urgency of the argument and the fascination which the starry sky has always had for mankind. It is worth using concepts that the general public can easily understand, so when dealing with night sky brightness a typical question is "How many stars can we see from our city?". The answer is not obvious. It depends on the observer's eye capability, atmospheric extinction and natural sky brightness, which depend in turn on the atmospheric conditions and the elevation of the observing site, the artificial sky background which depends on light pollution from the surrounding area and many other details. In this paper we present a method for obtaining maps of the number of stars visible in the hemisphere above to an average observer across large areas.

The availability of radiance-calibrated night-time satellite images of the Earth (Elvidge et al.

---
[*] falchi@istil.it

1999) allowed Falchi and Cinzano (Falchi 1998, Cinzano et al. 1999, Falchi & Cinzano 2000) to apply simple models for the propagation of light pollution to satellite data producing detailed maps of night sky brightness. Cinzano et al. (2000, 2001a, hereafter Paper 1 and Paper 2) used refined models based on Garstang to produce maps of zenith artificial night sky brightness in V and B bands (Paper 1 and Cinzano et al. 2001b), total night sky brightness at zenith and limiting magnitude (Paper 2), and hemispheric allsky maps at sites (Cinzano & Elvidge 2004). They accounted for Rayleigh scattering by molecules and Mie scattering by aerosols, extinction along light paths, Earth curvature, elevation, natural sky brightness, observed sky direction, eye capability, atmospheric extinction and in some cases mountain screening. Improvements on the modelling technique introduced by Cinzano et al. (2007) and Cinzano & Falchi (2012) provide a more general numerical solution for the radiative transfer problem applied to the propagation of light pollution in the atmosphere.

The lptran software package, developed by Cinzano (Cinzano et al. 2007; Cinzano & Falchi 2012), will allow, in future works, the production of an atlas of the number of visible stars taking into account (i) multiple scattering, (ii) wavelengths across all the visible spectrum, (iii) Earth curvature and its screening effects, (iv) elevation of sites and sources, (v) customized atmosphere (e.g. including thermal inversion layers), (vi) custom setup of different boundary layer aerosols and tropospheric aerosols, (vii) up to 5 aerosol layers in the upper atmosphere (e.g. fresh and aged volcanic dust and meteoric dust), (viii) variations of the scattering phase function with elevation, (ix) continuum and line gas absorption of many kinds, ozone included, (x) up to 5 cloud layers, (xi) wavelength dependent bidirectional reflectance of the ground surface (e.g. effect of snow), (xii) geographically variable upward light emission function. This modelling technique allows the production of global information on the average number of stars visible by an average observer from almost any site in the world. However, it would be incorrect to merely relate the limiting magnitude at the zenith to the number of stars visible in the entire sky and this merits a brief discussion. In fact, there is no fixed relation between the number of stars visible in the entire night sky hemisphere and the faintest stars visible at the zenith for several reasons: the number of stars of less than (i.e. brighter than) a given magnitude is not exactly exponential and somewhat depends on the chosen star catalogue; artificial and natural sky brightness change according to the direction of observation, requiring integration over the visible hemisphere or modelling a simple approximation; extinction and stars' apparent magnitude change with zenith distance.

In section 2 we describe the mapping technique and discuss it. In section 3 we present as an example a map of the number of stars visible to an average observer in Italy with a resolution of approximately 1 km and in section 4 we draw our conclusions.

## 2 Mapping technique

The number of visible stars depends on the limiting magnitude which in turn depends on the direction of observation due, in particular, to the effect of gradients of extinction and gradients of artificial and natural night-sky brightness. Then the total number of visible stars in a site is given by the integral of the density of visible stars per unit solid angle in the upward hemisphere:

$$n_{st} = \int_0^{2\pi} \int_0^{\pi/2} \rho_{st}(\theta,\phi) \sin\theta \, d\theta \, d\phi \tag{1}$$

where $\rho_{st}(\theta, \phi)$ is the number density of visible stars per unit of solid angle in the direction defined by zenith distance $\theta$ and azimuth $\phi$.

Here we summarize the steps to evaluate the number of visible stars $n_{st}$ in a site at the center of each land area covered by a pixel of a map:

(i) The artificial night sky brightness $b_{i,j,t}$ at the center of the land area $(i,j)$ in the direction of the line-of-sight defined by zenith distance $\theta_t$ and azimuth $\phi_t$ is (Cinzano & Falchi 2012; Paper 1):

$$b_{i,j,t} = \sum_h \sum_k e_{h,k} f_{a,u,r,w}(\theta_t, \phi_t, d_{i-h,j-k}, h_{i,j}, s_{h,k}) \ . \tag{2}$$

for each pair $(i,j)$ which define the positions of the observing site and each $(h,k)$, which define the position of the polluting area on the array. The index $t$ defines the direction of observation on the sky hemisphere. The observing site is assumed to be at the centre of each land area $(i,j)$ into which the territory has been divided. Here $e_{h,k}$ is the total upward flux emitted by the land area $(h,k)$, $f$ is the light pollution propagation library for the chosen atmosphere $a$, upward emission function $u$, ground surface function $r$, wavelength $w$, $d$ is the distance between the site and the source, $s$ is the source elevation and $h$ is the site elevation. The summations are extended to all the land areas around the site inside a distance for which their contributions are non-negligible.

The upward flux can be obtained from satellite data, such as those taken with the Operational Linescan System (OLS) carried by the U.S. Air Force Defense Meteorological Satellite Program (DMSP) satellites (Elvidge et al. 1999) or those taken with the Visible Infrared Imaging Radiometer Suite (VIIRS) Day/Night Band (DNB) using the composite maps of night-time lights produced from swath-level DNB data from the Earth Observation Group (EOG) at the National Oceanic and Atmospheric Administration's National Centers for Environmental Information (Elvidge et al. 2013). The normalized upward light emission function of each land area $(i,j)$ giving the relative intensity of the light emission at each elevation angle can be obtained from satellite radiance measurements (paper 1), retrieved from comparison with ground-based measurements (Falchi et al. 2016, Kocifaj 2017, Kocifaj et al. 2019), from airborne instrumentation (Kyba et al. 2013) or by modelling the average upward intensity of the lighting installations in the area as a function of the emission angle (Luginbuhl et al. 2009, Aubé & Roby 2014, Kocifaj 2018). The propagation function library can be obtained with lptran (Cinzano & Falchi 2012) or with models of propagation of light pollution in the atmosphere (Garstang 1984, 1986, 1987, 1988, 1989a, 1989b, 1989c, 1991a, 1991b, 1991c, 1992, 1993, 2000a; Cinzano 2000a, 2000b, 2000c; Aubé & Kocifaj 2012) accounting for Earth curvature, elevation, atmospheric conditions. Cinzano & Elvidge (2004) showed how to take into account mountain screening for zenith distances different from zero.

(ii) The natural night sky brightness $b_{\text{nat},i,j,t}$ at the centre of each land area $(i,j)$ in the direction of observation defined by $\theta_t, \phi_t$ can be obtained with Garstang (1989) models accounting for elevation and extinction and the input parameters can be tuned to fit measurements of sky background in unpolluted sites as described in Paper 2, sec. 2.2. Given that in some sites, but not in all, natural sky brightness is reported to be dependent on the solar cycle due to its effects on airglow, the night sky brightness at sea level should be corrected for the phase of the solar cycle or given for minimum (or average) solar activity.

(iii) The reader is referred to section 2.3 in Paper 2 for an extensive discussion and details, including the used values, on the content of this subsection. The limiting magnitude above the atmosphere $m_{i,j,t}$ at each grid point $(i,j)$ in the direction of observation defined by $\theta_t, \phi_t$ can be obtained from the total night sky brightness $b_{T\ i,j,t} = b_{i,j,t} + b_{nat,i,j,t}$ with the approach of Garstang (2000b) and Schaefer (1990,1993) accounting for eye capability and atmospheric extinction (Snell & Heiser 1968), as described in Paper 2, sec. 2.3:

$$m_{i,j,t} = c - 2.5 \log i'(b_{T\ i,j,t}) \tag{3}$$

where $i'$ is the illumination[2] produced over the atmosphere by a star at the threshold visibility for scotopic or photopic vision and $c$ is a constant given by Allen (1973 p. 197) for the units used by Garstang (2000b). The illumination $i'$ is related to the brightness of the night sky background $b$ (Garstang 2000b based on measurements of Blackwell (1946) and Knoll, Tousey and Hulburt(1946)):

$$i' = i'_1 i'_2 / (i'_1 + i'_2) \tag{4}$$
$$i'_1 = F_{a,1} F_{sc,1} F_{cs,1} F_{e,1} F_{s,1} \quad i_1(b_o, \theta) \tag{5}$$
$$i'_2 = F_{a,2} F_{sc,2} F_{cs,2} F_{e,2} F_{s,2} \quad i_2(b_o, \theta) \tag{6}$$
$$b_o = b / (F_a F_{sc} F_{cb}) \tag{7}$$

where $b_o$ is the observed background, $\theta$ is the stimulus size, i.e. the seeing disk diameter, $F_a$ takes into account ratio between average pupil area of the Knoll, Tousey, Hulburt (1946) and Blackwell (1946) observers and the pupil area of the assumed observer, $F_{sc}$ takes into account the Stiles-Crawford effect, $F_{cs}$ allows for the difference in color between the laboratory sources and the observed star, $F_e$ allows for star light extinction in the terrestrial atmosphere, taking into account that star magnitudes are given *outside the atmosphere*, $F_s$ allows for the acuity of any particular observer (defined so that $F_s < 1$ leads to a lower threshold $i$ and therefore implies an eye sensitivity higher than average due possibly to above average retinal sensitivity, observing experience, an above average eye pupil size or a better than average eye's point spread function (Navarro et al. 1997, 1998; Bara 2013), $F_{cb}$ allows for the difference in colour between the laboratory sources used in determining the relationships between $i'$ and $b_o$. The first equation was introduced by Garstang in order to put together smoothly the two functions $i'_1$ and $i'_2$, related respectively to the thresholds of scotopic and photopic vision, obtaining the best fit with cited measurements. The computation requires the definition of a number of parameters such as the observer's age, eye sensitivity, capability for averted vision, adaption of the eyes to the dark, observing with one or both eyes, the experience which makes an observer confident of a detection at a probability level different from the others (Schaefer 1990 reports a difference of 1 magnitude from 10% to 90% detection probability, corresponding respectively to the fainter suspected star and the fainter surely visible star), and the lengths of time for which the field has been observed. Schaefer 1990 reports roughly half a magnitude from 6 seconds to 60 seconds observation, but Crumey (2014) affirms, based on Mackworth (1948), that long exposure times degrade performance. Some differences could arise, for example, if we assume that our observer is really counting the stars or if (s)he is looking generically at the sky. An inexperienced observerwould probably count the star without properly using averted vision but when looking generically at the starry sky the main part of it would be seen from receptors in averted areas. So (s)he would be properly using averted vision without even realizing it. The colour index of the sky required for background colour correction could be inferred from maps of artificial night sky brightness in V and B bands.

---

[2] We use illumination and brightness for the flux of light per unit surface and light intensity per unit surface. In the particular case of the photopic sensitivity curve of the human eye, then they become illuminance and luminance respectively.

(iv) The limiting magnitude above the atmosphere is obtained by correcting for extinction $F_{e,1}$ and $F_{e,2}$ in eq. 5 and eq. 6. The vertical extinction in V band produced by molecules, aerosols and ozone is given by eq. 6 of Garstang (1991). A good Rayleigh airmass is given by Pickering (2002), the aerosol airmass by Rawlins (1992) as refined by Pickering (2002) and the ozone airmass by Schaefer (1990). We neglected extinction differences between V band and the visual band used in Garstang formulae. Extinction in the scotopic band can be estimated from the V band extinction (Garstang 2000b) or calculated from B and V extinctions.

(v) The number density $\rho_{i,j,t}$ of stars visible in a unit of solid angle of sky of given limiting magnitude $m_{i,j,t}$ above the atmosphere at the site $(i,j)$ can be obtained from statistical computations available in star catalogues. Figure 1 shows the total number of stars $\rho$ in the entire sky with apparent V magnitude above the atmosphere lessthan (i.e. stars brighter than) $m$, as obtained from the statistics published in the main catalog of luminous stars, the SAO Star Catalog (1966, p.XIX), the Bright Stars Catalogue (Hoffleit 1964), the Sky Catalogue 2000 (Hirshfeld, Sinnott 1982, tab. II), the Tycho Catalogue (Sinnott, Perryman 2000, p.VII) and Seares et al. (1929, tab. XVIII) as revised by (Allen 1973, p.244). Differences between V band magnitudes of Tycho and Tycho2 catalogues are negligible for our purposes (Hog et al. 2000, fig. 5). We also neglected differences between visual and V band magnitudes. For some catalogs we had to convert the number of stars in the range $m - 1/2$, $m + 1/2$ in the total number of stars of magnitude brighter than $m$. The best fitting line in figure 1 is a good approximation for our purposes.

The average density of stars per unit of solid angle $\rho_{i,j,t}$ which is visible at the site $(i,j)$ in the direction of observation defined by $\theta_t, \phi_t$ when the limiting magnitude above the atmosphere $m^*_{i,j,t}$ can be obtained dividing the best fitting function of figure 1 for $4\pi$ steradian:

$$\rho_{i,j,t} = 0.34682 \ e^{1.169 \ m^*_{i,j,t}} \tag{8}$$

(vi) The total number of stars visible in the upper hemisphere from an observer at the centre of the land area $n(i,j)$ can be computed from $\rho_{i,j,t}$ following eq. 1. The integral can be replaced by the numerical sum: $n_{i,j} = \sum_t q_t \rho_{i,j,t}$ where $q_t$ are integration coefficients chosen on the basis of the numerical integration scheme adopted. This numerical integration is not trivial. Computational times should be balanced with a number of points in the sky hemisphere sufficiently large to guarantee a good integration. If there is some a priori information regarding the functional form of $\rho_{i,j,t}$ the integration times can be strongly reduced by using appropriate cubatures e.g. those based on Albrecht's grids (Bará et al 1996; Ríos et al 1997) or other optimal sampling patterns (Ramos-López et al, 2016). An example of numerical summation over 25 areas in the sky hemisphere is shown in Figure 2.

It is worth pointing out that in general the number of stars visible in the sky is always lower, sometimes substantially, to the number of stars above the horizon brighter than the limiting magnitude at zenith. In fact, the density of visible stars diminishes going down to the horizon due to increasing atmospheric extinction and, in urbanized areas, to increasing light pollution. As an example, the first 20 degrees of sky above the horizon occupy about one third of the solid angle of the entire hemisphere, i.e. in the first 20 degrees are found about one third of theoretically visible stars. Stars at 10° above the horizon experience a magnitude extinction of about six times higher than at the zenith and are hidden by a usually much brighter background due to light pollution.

## 3 Results

In Figure 3 we present as an example the simplified map of Italy showing the number of stars visible to an average observer. The simplification arises from the fact that in densely populated areas, where at each site the artificial sky brightness is produced by a multitude of different sources, the night sky brightness at different zenith distances can be inferred from the zenith brightness, so that a computation of the real brightness at different azimuths and altitude is not mandatory. In the map, the levels from white to black correspond to 0-200, 200-400, 400-700, 700-1000, 1000-1300, >1300 stars. It is worth noting that the highest number of visible stars is achievable only with very low light pollution levels and high transparency. In fact, at sea level, even with no pollution at all, it is not possible to see the highest number of stars. The map was computed with Garstang libraries of light pollution propagation for a clean atmosphere with aerosol clarity $K=1$, corresponding to a vertical extinction in the V band of $\Delta m = 0.33$ mag at sea level, $\Delta m = 0.21$ mag at 1000m o.s.l., $\Delta m = 0.15$ mag at 2000m o.s.l., horizontal visibility at sea level $\Delta x = 26$ km, optical depth $\tau = 0.3$ so double scattering approximation is adequate The number of visible stars is computed for observers of average experience and capability $F_s = 1$, aged 40 years, with eyes adapted to the dark, observing with both eyes. Input radiance data were obtained from the DMSP composite maps of night-time lights produced from the Earth Observation Group (EOG) at the National Oceanic and Atmospheric Administration's National Centers for Environmental Information. Country boundaries are approximate. Original maps are 4800×4800 pixel images saved in 16-bit standard fits format with fitsio Fortran-77 routines developed by HEASARC at the NASA/GSFC. Each pixel is 30 x 30 arcsec$^2$ in size in longitude/latitude projection. We refer the reader to the papers cited in sec.2 for a discussion of details of the computation of artificial night sky brightness and adopted assumptions. Main ones are: artificial night sky brightness is computed integrating contribution by sources inside a circle of about 200 km around the site, site elevation is accounted in the range 0-3000 m (higher ones are set to 3000 m because the map is intended to evaluate the number of visible stars in urbanized areas rather than in observatory sites), sources are assumed at sea level. The library adopts Garstang's average upward emission function (Paper 1) for night-time lighting of urbanized areas so map accuracy is uncertain near particular light pollution sources like sea platforms, mines, fishing fleets. Mountain screening and snow reflection have been neglected but can be accounted in further works.

This work presents a method to calculate the number of visible stars as seen by an average observer, so a direct comparison with observations is beyond its scope. Nonetheless, all the steps involved in the computation (except the very last one, counting the stars by a great number of observers and from a great number of sites) have already been validated in previous works. The sky brightness in every direction in the sky has been validated mainly by Cinzano and Elvidge (2004); the limiting magnitude dependence on sky brightness has been explored using the models of Garstang (2000b), Schaefer (1990,1993) and by Cinzano et al. (2001a); the natural sky brightness by Garstang (1989); the number of stars for a given limiting magnitude in this work, using the results of common star catalogs; the atmospheric extinction by Garstang (1991, 2000b).

A comparison with observations will depend on the availability of observations of the number of visible stars from several sites, or, at least, the limiting magnitude at different azimuths and zenith distances. Of course, a direct measure of the number of visible stars will not be possible, except for the very polluted skies visible from large urban areas where the number of stars can be directly counted. In most places an estimate of the number should be computed from the count of stars inside relatively small solid-angle portions of the sky in appropriate azimuth and altitude directions. For example, a square frame of 10 cm by 10 cm aperture, put at 40 cm distance from the eyes, shows about 1/90 of the sky. Pointing towards the zenith, at 45° above the horizon toward cardinal points and at 15° above the horizon at halfway between the cardinal points, summing the counts and multiplying by 10 will give an estimation of the number of stars visible (e.g. in Falchi 2009). We hope that this work may help to stimulate observation campaigns to obtain the great number of observations needed for a direct comparison with the prediction of the model.

## 4 Conclusions

In this paper we present a method for computing the number of visible stars in the night sky hemisphere, taking into account the main factors, including the number of visible stars brighter than a given magnitude obtained from the main star catalogs, the altitude of the observing site, the atmospheric extinction in the different zenith distances, the artificial and natural sky brightness in the different directions in the sky, and the observer's visual acuity and experience. The method can be applied to radiance-calibrated satellite data in order to compute maps of the number of stars visible from large areas. We also present a simplified sample map for a portion of Europe. In future, this method can be used to produce an atlas of the number of visible stars from every site in the world.

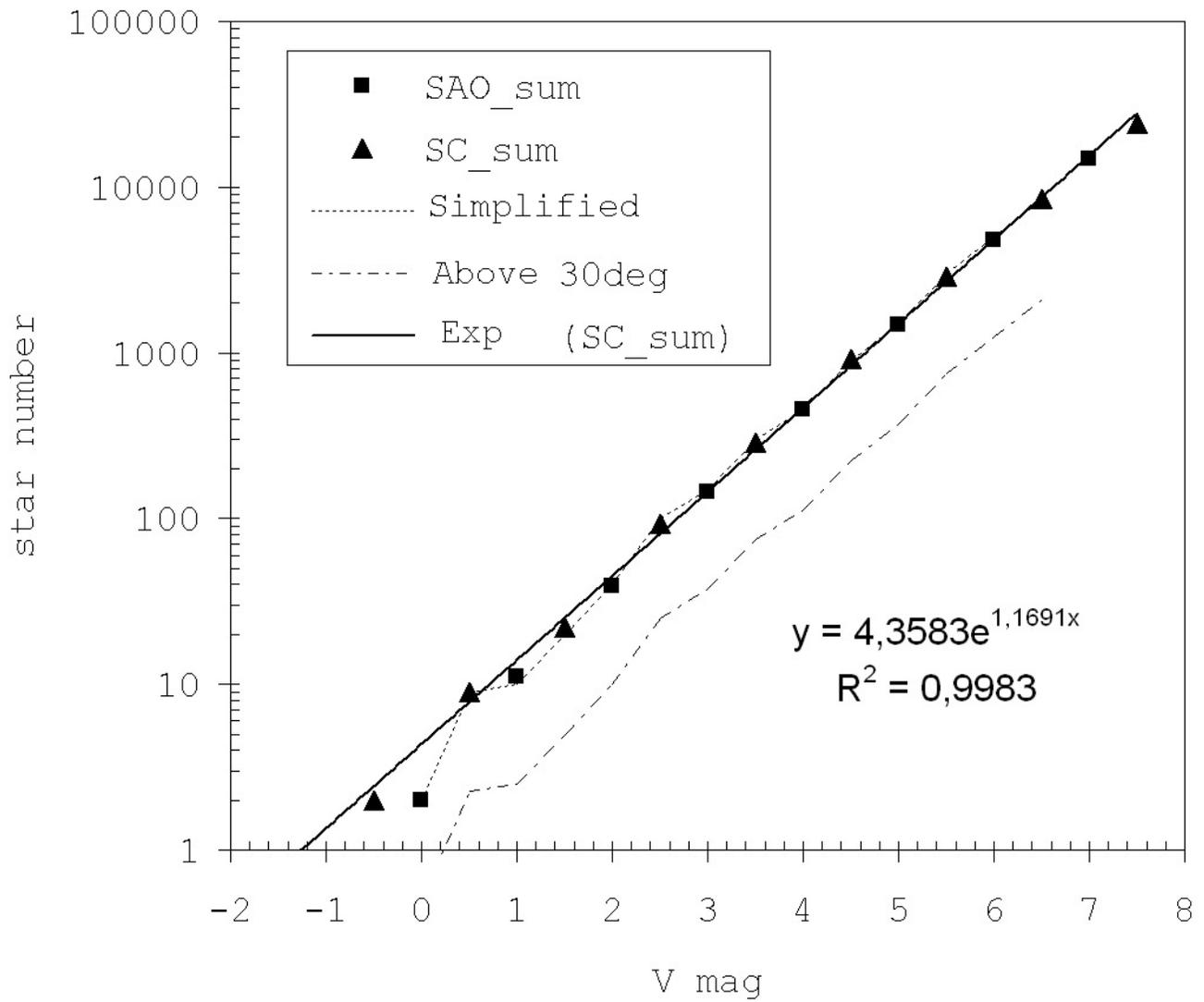

*Figure 1: Total number of stars in the entire sky sphere brighter than a given magnitude (above the atmosphere). The sum of the number of stars brighter than a given magnitude is obtained from the SAO Star Catalog (squares) and Sky Catalogue 2000 (triangles). The line-dotted line shows the number of stars above 30 degrees of altitude above the horizon.*

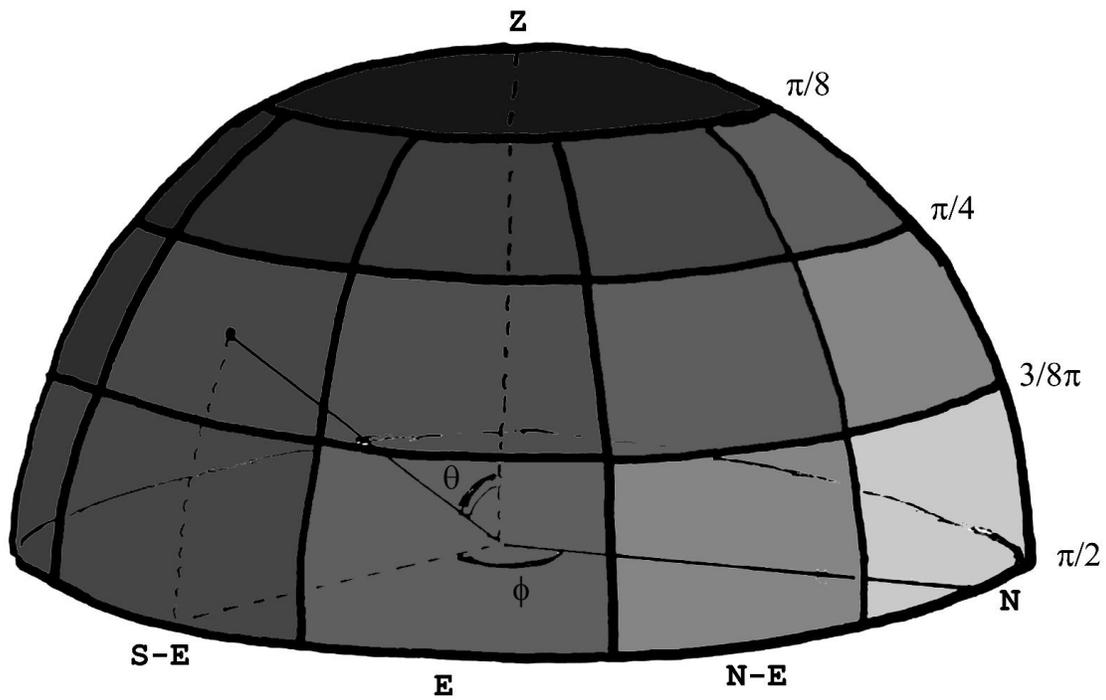

Figure 2. Example of numerical summation where the night hemisphere is divided into 25 areas. In this case the area surrounding the zenith Z has a $q_t$ =0.478 sr, the other rectangular sectors progressively approaching the horizon are 0.170 sr, 0.255 sr and 0.301 sr. The shades of gray indicate qualitatively the visible star density $\rho_{st}(\theta, \phi)$. In this example the main source of pollution is towards the North.

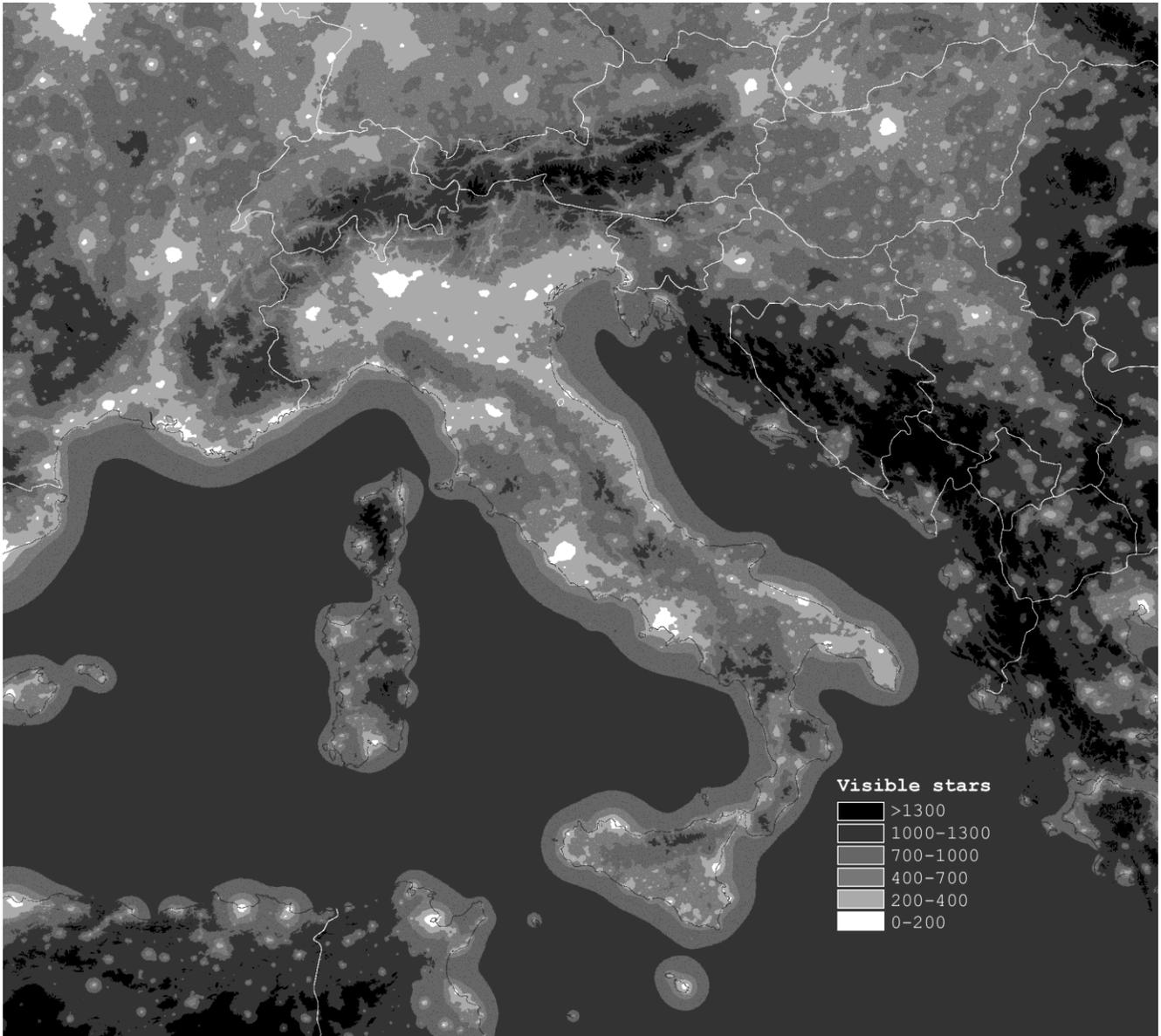

*Figure 3: Number of stars visible to an average observer in Italy on standard clear nights in the upward hemisphere.*

## Acknowledgments

We are indebted to Prof. Roy Garstang of JILA-University of Colorado for having paved the way towards modern research on light pollution. We dedicate this work in memory of him, our friend and mentor. We thank Salvador Bará for his critical reading and for his useful suggestions and Bob Mizon for his help in English language. We acknowledge the EROS Data Center, Sioux Falls, USA for kindly providing us with their GTOPO30 digital elevation model. Part of this work has been supported by the University of Padua CPDG023488 and the Italian Space Agency I/R/160/02.